\documentclass[12pt]{article}

\usepackage{graphicx}
\usepackage{amsfonts}
\usepackage{amsthm}
\usepackage{amsmath}
\usepackage[top=3cm, bottom=3.25cm, left=2.75cm, right=2.75cm]{geometry}

\newcommand{\beq}{\begin{equation}}
\newcommand{\eeq}[1]{\label{#1}\end{equation}}
\newcommand{\ber}{\begin{eqnarray}}
\newcommand{\eer}[1]{\label{#1}\end{eqnarray}}
\newcommand{\re}[1]{(\ref{#1})}

\numberwithin{equation}{section}

\newcommand{\nZZ}{N\!=\!(2,2)}

\newcommand{\bbD}[1]{\mathbb{D}_{#1}}
\newcommand{\bbDB}[1]{\bar{\mathbb{D}}_{#1}}
\newcommand{\bbX}[1]{\mathbb{X}^{#1}}

\newcommand{\+}{{\scriptscriptstyle{+\!\! +}}}
\newcommand{\pp}{{\pm\!\!\pm}}

\newcommand{\ep}[1]{\epsilon^{#1}}
\newcommand{\epb}[1]{\bar{\epsilon}^{\,#1}}

\newcommand{\Jp}{J^{\mbox{\tiny$(+)$}}}
\newcommand{\Jm}{J^{\mbox{\tiny$(-)$}}}
\newcommand{\Jpm}{J^{\mbox{\tiny$(\pm)$}}}
\newcommand{\Up}{U^{\mbox{\tiny$(+)$}}}
\newcommand{\Um}{U^{\mbox{\tiny$(-)$}}}
\newcommand{\Vp}{V^{\mbox{\tiny$(+)$}}}
\newcommand{\Vm}{V^{\mbox{\tiny$(-)$}}}
\newcommand{\Upm}{U^{\mbox{\tiny$(\pm)$}}}
\newcommand{\Vpm}{V^{\mbox{\tiny$(\pm)$}}}

\newcommand{\pip}{\pi^{\mbox{\tiny$(+)$}}}

\def\one{{1\!\! 1}}
\def\+{{+\!\!\!+}}

\newcommand{\nn}{\nonumber}

\newcommand{\pa}[1]{\partial_{#1}}

\newcommand{\eg}{\textit{e.g.},~}
\newcommand{\half}{\textstyle{\frac 1 2}}

\begin{document}
\renewcommand{\theequation}{\thesection.\arabic{equation}}
\setcounter{page}{0}
\thispagestyle{empty}
\begin{flushright} \small

\end{flushright}
\renewcommand{\theequation}{\thesection.\arabic{equation}}
\setcounter{page}{0}
\thispagestyle{empty}
\begin{flushright} \small
UUITP-14/12 \\
YITP-SB-12-23\\
\end{flushright}

\smallskip
\begin{center}
 \LARGE
{\bf  Semichiral Sigma Models with 4D Hyperk\"ahler Geometry}
\\[12mm]
 \normalsize
{\bf M.~G\"oteman$^{a}$, U.~Lindstr\"om$^{a}$ and M.~Ro\v cek$^{b}$}, \\[8mm]
 {\small\it
$^{a}$Department Physics and Astronomy,\\
Division for Theoretical Physics,\\
Uppsala University, \\ Box 803, SE-751 08 Uppsala, Sweden\\
~\\
$^{b}$C.N.Yang Institute for Theoretical Physics,\\ Stony Brook University, \\
Stony Brook, NY 11794-3840,USA\\}
\end{center}
\vspace{10mm} \centerline{\bfseries Abstract} \bigskip

\noindent 
Semichiral sigma models with a four-dimensional target space do not support extended $N=(4,4)$ supersymmetries off-shell  \cite{Goteman:2009xb},  \cite{Goteman:2009ye}. We contribute towards the understanding of the non-manifest on-shell transformations in $(2,2)$ superspace by analyzing the extended on-shell supersymmetry of such models and find that a rather general ansatz for the additional supersymmetry (not involving central charge transformations) leads to hyperk\"ahler geometry. We give non-trivial examples of these models.
\eject

%======================================================================
%======================================================================
\tableofcontents
\section{Introduction}

In a previous paper  \cite{Goteman:2009ye}, we presented the general structure of semichiral sigma models with $(4,4)$ supersymmetry. We found conditions for invariance of the action  and interesting geometric structures related to simultaneous integrability  (Magri-Morosi concomitants) and a weaker conditions than (almost) complex structures for the transformation matrices  (Yano f-structures).
This rich mathematical context prompt us to take a closer look at  specific models. 

In \cite{Goteman:2009ye} we treated both  off-shell and on-shell supersymmetric $(4,4)$ models with manifest $(2,2)$ supersymmetry. One particular model where the  non-manifest  supersymmetry can only close on-shell, is the case of one left and one right semichiral field corresponding to a four-dimensional target space.\footnote{The off-shell $N=(4,4)$ {\em pseudo}-supersymmetry for a semichiral sigma model with four-dimensional target space was discussed in detail in \cite{Goteman:2009xb}.} These models are simple enough that a lot of the calculations can be  carried out explicitly. They also enjoy a number of special properties such as carrying an almost (pseudo-) hyperk\"ahler structure and having a $B$-field that is governed by a single function.

In the present paper we start from the same general ansatz for the extra supersymmetries as in \cite{Goteman:2009ye}, we then solve the conditions for invariance of the action and discover that on-shell closure of the algebra follows from these, with one additional input from the algebra.  The solution leads to a  geometry which  is necessarily hyperk\"ahler.  Note that we are not proving that $(4,4)$ supersymmetry in four dimensions restricts the target space geometry to be hyperk\"ahler, since there may be more general ans\"atze combining supersymmetry with central charge transformations. We briefly discuss this option in our conclusions.

We relate our solution to geometric conditions from \cite{Goteman:2009ye} and illustrate our findings in an explicit (non-trivial) example.

The paper is organized as follows. Section \ref{Semichiral sigma models} contains background material, definitions and sets the notation for the paper. In section \ref{On-shell $N=(4,4)$ supersymmetry} we give the derivation of our conditions for invariance of the action and on-shell closure while section \ref{HK} examplifies them. Section \ref{summary} contains our conclusions and in the appendix we have collected the special case of linear transformations, the explicit form of the complex structures and the metric for the example in section \ref{HK}.
%======================================================================
%======================================================================
\section{Semichiral sigma models}
\label{Semichiral sigma models}
Consider a generalized K\"ahler potential with one left- and one right semichiral field and their complex conjugates, $K(\bbX{L}, \bbX{R})$, where $L=(\ell, \bar \ell)$ and $R=(r, \bar r)$. The action,
\beq
	S=\int d^2x d^2\theta d^2\bar{\theta}K(\bbX{L}, \bbX{R})
\eeq{action}
has manifest $\nZZ$ supersymmetry. The supersymmetry algebra is defined in terms of the anti-commutator of the covariant supersymmetry derivatives as
\beq
\{\bbD{\pm},\bbDB{\pm}\}=i\pa{\pp}
\eeq{susyalgebra}
and the semichiral fields are defined by their chirality constraints as
\beq
	\bbDB{+}\bbX{\ell} = 0~, \quad \bbDB{-} \bbX{r} = 0~.
\eeq{defining_semichirals}

The geometry of the model is governed by two complex structures $\Jp$ and $\Jm$ that both preserve the metric $g$
\ber
\Jpm{}^{t}g\Jpm =g
\eer{gpres}
as well as by an anti-symmetric $B$-field whose field strength $H$ enters in the form of torsion in the integrability conditions
\beq
0=\nabla^{(\pm)}\Jpm=\left(\partial+\Gamma^{(0)}\pm\half Hg^{-1}\right)\Jpm~,
\eeq{int}
where $\Gamma^{(0)}$ is the Levi-Civita connection. These conditions identify the geometry as bi-hermitean \cite{Gates:1984nk}, or generalized K\"ahler geometry (GKG) \cite{Gualtieri:2003dx}. 

The fact that our superfields are semichiral specifies the  GKG as being of symplectic type where the metric $g$ and the $B$-field take the form\footnote{This gives the $B$ field in a particular global gauge as $B=B^{(2,0)}+B^{(0,2)}$ with respect to both complex structures.}
\ber\nn
g&=&\Omega [\Jp, \Jm]\\
B&=&\Omega \{\Jp, \Jm \}~.
\eer{coms}
The matrix $\Omega$ is defined as 
\beq
\Omega=\half\left(\begin{array}{cc}
0&K_{LR}\cr
-K_{RL}&0\end{array}\right)
\eeq{omega}
and the submatrix $K_{LR}$ is the Hessian
\beq
K_{LR}=\left( \begin{array}{cc}K_{\ell r}&K_{\ell\bar r}\cr
K_{\bar \ell r} &K_{\bar \ell\bar r}\end{array}\right)~.
\eeq{103}

An additional condition results from the target space being four-dimensional and reads  \cite{Buscher:1987uw}
\beq
 \{\Jp, \Jm\}=2 c \one,~~~~\Rightarrow B=2 c \,\Omega~,
\eeq{fourd}
where  $c= c(\bbX{L},\bbX{R})$. For reference, we rewrite this relation as
\beq
 (1- c)|K_{\ell \bar r}|^2+(1+ c)|K_{\ell r}|^2=2K_{\ell\bar\ell}K_{r\bar r}~.
 \eeq{106}
The condition \re{fourd} allows us to construct an $SU(2)$ worth of almost (pseudo-) complex structures\footnote{In higher dimensions than four, almost hyperk\"ahler implies hyperk\"ahler \cite{Crichigno:2011aa}.} $J^{(1)},J^{(2)},J^{(3)}$, \cite{Goteman:2009xb}, \cite{Buscher:1987uw}, \cite{Lindstrom:2005zr},
\ber\nn
J^{(1)}&:=&\frac {1}{\sqrt{1-c^2}}\left(\Jm+ c\Jp \right)~,\nn\\
J^{(2)}&:=&\frac 1 {2\sqrt{1-c^2}}[\Jp, \Jm ]~,\nn\\
J^{(3)}&:=&\Jp{}~.
\eer{su2}
For $|c|<1$ the geometry is almost hyperk\"ahler, while for $|c|>1$ the geometry is almost pseudo-hyperk\"ahler \cite{Goteman:2009xb}.

%======================================================================
%======================================================================
\section{On-shell $N=(4,4)$ supersymmetry}
\label{On-shell $N=(4,4)$ supersymmetry}
\subsection{The ansatz}
In this first subsection we recapitulate some definitions from \cite{Goteman:2009ye}. 

Additional supersymmetry transformations on the semichiral fields must preserve the chirality constraints \re{defining_semichirals}. We make the following ansatz for the additional supersymmetry,
\ber\nn
\delta \bbX{\ell} &=&\bar\epsilon^+ \bbDB{+}f(\bbX{L}, \bbX{R}) + g(\bbX{\ell})\bar\epsilon^{-}\bbDB{-} \bbX{\ell}\, + h(\bbX{\ell})\epsilon ^- \bbD{-}\bbX{\ell}~,\\
\delta \bbX{\bar\ell} &=& \ep{+}\bbD{+} \bar f(\bbX{L}, \bbX{R}) + \bar g(\bbX{\bar \ell})\ep{-}\bbD{-}\bbX{\bar\ell}+\bar h(\bbX{\bar \ell}) \epb{-}\bbDB{-}\bbX{\bar \ell}~,\nn\\
\delta \bbX{r}&=& \bar\epsilon^- \bbDB{-}\tilde f(\bbX{L}, \bbX{R}) +\tilde g(\bbX{r})\bar\epsilon^+ \bbDB{+}\bbX{r}+\tilde h(\bbX{r})\epsilon^+ \bbD{+}\bbX{r}~,\nn\\
\delta\bbX{\bar r} &=& \ep{-}\bbD{-} \bar{\tilde f}(\bbX{L}, \bbX{R}) + \bar{\tilde g}(\bbX{\bar r})\ep{+}\bbD{+}\bbX{\bar r}+\bar{\tilde h}(\bbX{\bar r}) \epb{+}\bbDB{+}\bbX{\bar r}~.
\eer{defining_ansatz_2}
For later convenience, a compact form of these transformations will be useful. We thus introduce transformation matrices
\beq
	\begin{array}{lcllcl}
	U^{(+)} &=& \left(\begin{array}{cccc}
	\ast & f_{\bar\ell} & f_r & f_{\bar r}\\
	\ast & 0 & 0 & 0\\
	\ast & 0 & \tilde g & 0\\
	\ast & 0 & 0 & \bar{\tilde h} 
	\end{array}\right), \quad &
	V^{(+)} &=& \left(\begin{array}{cccc}
	0 & \ast & 0 & 0\\
	\bar{f}_\ell & \ast & \bar{f}_{\bar r} & \bar{f}_r\\
	0 & \ast &  \tilde h & 0\\
	0 & \ast & 0 & \bar{\tilde g} 
	\end{array}\right),\\
	&&&\\
	U^{(-)} &=& \left(\begin{array}{cccc}
	g & 0 & \ast & 0\\
	0 & \bar h & \ast & 0\\
	\tilde{f}_\ell & \tilde{f}_{\bar \ell} & \ast & \tilde{f}_{\bar r}\\
	0 & 0&\ast & 0
	\end{array}\right), \quad &
	V^{(-)} &=& \left(\begin{array}{cccc}
	h & 0 &0 & \ast \\
	0 & \bar g &0 & \ast \\
	0 & 0 & 0 & \ast\\
	\bar{\tilde f}_{\ell} & \bar{\tilde f}_{\bar \ell} & \bar{\tilde f}_r & \ast\\
	\end{array}\right).
\end{array}
\eeq{transformation matrices}
One column in each of the matrices is arbitrary. Further writing the semichiral fields as a vector $\bbX{i}$ where $i=(\ell, \bar\ell, r, r)$, 
the transformations  read
\beq
\delta \bbX{i} = \epb{\alpha} {U^{\mbox{\tiny$(\alpha)$}}} {}^i_j\bbDB{\alpha}\bbX{j} +  \ep{\alpha} {V^{\mbox{\tiny$(\alpha)$}}} {}^i_j\bbD{\alpha}\bbX{j}~,  
\eeq{ansatz_compact}
where the spinor index $\alpha$ takes the values  $+$ and $-$.
%======================================================================
\subsection{No off-shell supersymmetry}
\label{No off-shell supersymmetry}
Consider supersymmetry closure of the transformations in \re{defining_ansatz_2}. The supersymmetry algebra defined in \re{susyalgebra} requires that two subsequent transformations commute to a translation, \eg that $[\delta^{(+)}_1, \delta^{(+)}_2]\bbX{\ell}=i\ep{+}_{[2}\epb{+}_{1]}\pa{}\bbX{\ell}$. But the transformations in \re{defining_ansatz_2} commute to
\ber\nn
[\delta_1, \delta_2]\bbX{\ell}
&=&-\epsilon^+_{[2}\bar\epsilon^+_{1]}\left(|f_{\bar\ell} |^2 i\pa{\+}\bbX{\ell}+(f_{\bar \ell}\bar f_r+f_r\tilde h) \bbDB{+} \bbD{+}\bbX{r}+(f_{\bar \ell}\bar f_{\bar r}+f_{\bar r}\bar{\tilde g}) \bbDB{+} \bbD{+}\bbX{\bar r}\right)\nn\\
&& + \bar\epsilon^-_{[2}\epsilon^-_{1]}(-gh) i\partial_= \bbX{\ell}+\dots~,
\eer{no_offshell}
where the dots represent the mixed $\ep{+}\ep{-}$-terms in the algebra. 
Since $|f_\ell|^2\neq -1$,  the $\ep{+}\epb{+}$-part of the algebra \re{defining_ansatz_2} can never close off-shell\footnote{From the derivation we see that the full statement is that we cannot have a left (or right) supersymmetry off-shell.}. In section \ref{On-shell algebra closure}, we will see that the algebra closes on-shell.

We do not get a contradiction from the $\ep{-}\epb{-}$-part of the algebra, however;  it closes  if and only if
\beq
g h=-1~.
\eeq{6}

%======================================================================
\subsection{Invariance of action}

The action in \re{action} is invariant under the $\bar\delta{}^{(+)}$-transformations if and only if  the Lagrangian satisfies the following partial differential equations \cite{Goteman:2009ye}
\beq
(K_i \Up{}^i_{[j})_{k]}\bbDB{+}\bbX{j}\bbDB{+}\bbX{k} = 0~,
\eeq{PDEs_summary}
together with the corresponding equations for the other transformation matrices.
Explicitly
\ber
f_r K_{\ell\bar\ell} - f_{\bar \ell} K_{\ell r} + \tilde g K_{\bar\ell r} &=& 0~,\nn\\
f_{\bar r}K_{\ell\bar\ell}-f_{\bar\ell} K_{\ell \bar r}+\bar {\tilde h}K_{\bar \ell\bar r}&=&0~,\nn\\
(\tilde g-\bar{\tilde h}) K_{r\bar r} - f_{\bar r} K_{r\ell} + f_{r} K_{\bar r \ell} &=& 0~,
\eer{PDEs_original_form}
and 
\ber
\tilde f_\ell K_{r\bar r} - \tilde{f}_{\bar r} K_{r\ell} + g K_{\bar r \ell} &=& 0~,\nn\\
\tilde f_{\bar\ell} K_{r\bar r} + \bar h K_{\bar r\bar \ell} -{\tilde f_{\bar r}} K_{r \bar \ell} &=& 0~,\nn\\
(g-\bar h) K_{\ell\bar\ell} - \tilde{f}_{\bar\ell} K_{\ell r} + \tilde f_{\ell} K_{\bar\ell r} &=& 0~.
\eer{PDEs_original_form_tilde}
In addition we have the relations complex conjugate to those in \re{PDEs_original_form} and \re{PDEs_original_form_tilde} and 
a derived useful relation between $\tilde h$ and $\tilde g$;
 \beq
\frac{\bar h}{g} = \frac{\bar{\tilde h}}{\tilde g} =  \frac{K_{\ell\bar\ell} K_{r\bar r}-|K_{\ell \bar r}|^2}{K_{\ell\bar\ell} K_{r\bar r}-|K_{\ell r}|^2}~.
\eeq{g_h_information}

\subsection{Integrability}

In \cite{Goteman:2009ye} we discuss on-shell closure of the $(4,4)$ algebra in terms of the $SU(2)$ set of complex structures $J^{(A)}_{(\pm)}$ known to exist from the $(1,1)$ reduction. We show that all the $(2,2)$ closure conditions are satisfied on-shell by relating them to expressions involving the $J^{(A)}_{(\pm)}$s.
In other words, we prove that the $(2,2)$ conditions needed for additional supersymmetry of the $(2,2)$ model are equivalent to the $(1,1)$ conditions 
needed for extra (on-shell) supersymmetry of the corresponding $(1,1)$ model.

Here we follow a different route. We assume that the systems of linear partial differential equations \re{PDEs_original_form} and \re{PDEs_original_form_tilde} are integrable and show that this, together with $g,\tilde g, h$ and $\tilde h$ all being constant, is sufficient to prove 
that all the $(2,2)$ closure conditions are satisfied on-shell.

Integrability of \re{PDEs_original_form} and \re{PDEs_original_form_tilde} in turn, may be discussed in terms of the usual machinery for analyzing systems of linear partial differential equations. We do not include such an analysis, but solve the equations in examples below.

A final comment on the relation to the analysis in  \cite{Goteman:2009ye} is relegated to Appendix B. There we show the  $(2,2)$ relation
\beq
\half \alpha~\!\left(J^{(1)}_{(\pm)}-iJ^{(2)}_{(\pm)}\right)=U^{(\pm)}\pi^{(\pm)}~,
\eeq{onshellcndn}
found in  \cite{Goteman:2009ye} and relating the abstract $J^{(A)}$s to the transformation matrices \re{transformation matrices}, indeed holds for the two
additional complex structures  $J^{(1)}_{(+)}$ and  $J^{(2)}_{(+)}$ explicitly constructed in \re{su2}.

\subsection{On-shell algebra closure}
\label{On-shell algebra closure}

Before discussing how to close the algebra on-shell, we point to some consequences of imposing the full algebra.

In \re{6}, we have seen that the $(-)$-part of the algebra for $\bbX{\ell}$ closes  if and only if $gh=-1$. The same is true for the $(+)$-part of the algebra for $\bbX{r}$; it closes  if and only if $\tilde g\tilde h=-1$. From \re{g_h_information} we then deduce that
\beq
\hbox{det} (K_{LR}) \ne 0~.
\eeq{determinant_related_to_bc}
This is a familiar condition which, amongst other things, ensures that a non-degenerate geometry can be extracted from the action \cite{Lindstrom:2005zr}. 
A further consequence of  $\tilde g\tilde h=-1=gh$ in conjunction with \re{PDEs_original_form} and 
\re{PDEs_original_form_tilde}  is that \re{106} is satisfied with
\beq
c = \frac{1-|g|^2}{1+|g|^2}=\frac{1-|\tilde g|^2}{1+|\tilde g|^2}~.
\eeq{hat_c_in_g}
Since we have assumed that $g= g(\bbX{\ell})$ and $\tilde g= \tilde g(\bbX{r})$, the relations \re{hat_c_in_g} tell us that $g, \tilde g, h$ and $\tilde h$ are all constant.

We now turn to the on-shell closure.
Two subsequent transformations defined in \re{defining_ansatz_2} acting on a left-semichiral field commute to
\ber
[\delta_1, \delta_2]\bbX{\ell} &=&
 \bar\epsilon^+_{[2}\epsilon^+_{1]} \left[\mathcal{M}(\Up, \Vp)^\ell_{jk}\bbDB{+}\bbX{j}\bbD{+}\bbX{k}-(\Up\Vp)^{\ell}_j\bbDB{+}\bbD{+}\bbX{j}-(\Vp\Up)^{\ell}_j\bbD{+}\bbDB{+}\bbX{j}\right]\nn\\
&+&\bar\epsilon^-_{[2}\epsilon^-_{1]} \left[\mathcal{M}(\Um, \Vm)^\ell_{jk}\bbDB{-}\bbX{j}\bbD{-}\bbX{k}-(\Um\Vm)^{\ell}_j\bbDB{-}\bbD{-}\bbX{j}-(\Vm\Um)^{\ell}_j\bbD{-}\bbDB{-}\bbX{j} \right]  \nn\\
&+& \bar\epsilon^+_{[2}\bar\epsilon^-_{1]} \left[\mathcal{M}(\Up, \Um)^\ell_{jk}\bbDB{+}\bbX{j}\bbDB{-}\bbX{k}-[\Um,\Um]^{\ell}_j\bbDB{+}\bbDB{-}\bbX{j}\right]\nn\\
&+& \bar\epsilon^+_{[2}\epsilon^-_{1]}\left[\mathcal{M}(\Up, \Vm)^\ell_{jk}\bbDB{+}\bbX{j}\bbD{-}\bbX{k}-[\Up,\Vm]^{\ell}_j\bbDB{+}\bbD{-}\bbX{j}\right]\nn\\
&+& \epb{-}_{[2}\epb{-}_{1]}\left[\mathcal{N}(\Um)^\ell_{jk}\bbDB{-}\bbX{j}\bbDB{-}\bbX{k}\right]
+\ep{-}_{[2}\ep{-}_{1]}\left[\mathcal{N}(\Vm)^\ell_{jk}\bbD{-}\bbX{j}\bbD{-}\bbX{k}\right]\nn\\
&+&
 \epb{+}_{[2}\epb{+}_{1]}\left[\mathcal{N}(\Up)^\ell_{jk}\bbDB{+}\bbX{j}\bbDB{+}\bbX{k}\right],
\eer{closure_of_XL_geometry}
where the transformation matrices $\Upm$ and $\Vpm$ are defined in \re{transformation matrices}. The Nijenhuis tensor $\mathcal{N}$ and the Magri-Morosi concomitant $\mathcal{M}$ are defined as 
\ber
\mathcal{N}(I)^i_{jk} &=& I^l_{[j}I^i_{k],l}+I^i_l I^l_{[j,k]}~,\nn\\ 
\mathcal{M}(I,J)^i_{jk} &=& I^l_{j}J^i_{k,l}-J^l_{k}I^i_{j,l}+J^i_lI^l_{j,k}-I^i_l J^l_{k,j}~.
\eer{nijenhuis_and_magri}
We will now show that each of the terms in \re{closure_of_XL_geometry} close to a supersymmetry algebra on-shell and discuss the geometric interpretation.
In section \ref{No off-shell supersymmetry} we have seen that the transformations in \re{defining_ansatz_2} cannot close to a supersymmetry off-shell. Hence we have to go on-shell. The field equations that follow from the action in \re{action} are
\ber\nn
\bbDB{+} K_\ell &=&0~,\\
\bbDB{-}K_r&=&0~,
\eer{7}
and their complex conjugates. To investigate on-shell closure, we use the the first equation to solve  for, \eg $\bbDB{+}\bbX{\bar\ell}$:
\beq
\bbDB{+}\bbX{\bar\ell}=-\frac{1}{K_{\ell\bar \ell}}(K_{\ell r}\bbDB{+}\bbX{r}+K_{\ell \bar r}\bbDB{+}\bbX{\bar r})~.
\eeq{8}
Using the expressions for the transformation functions in \re{PDEs_original_form}-\re{g_h_information} and the on-shell relation \re{8}, the last term in the algebra for $\bbX{\ell}$, \re{closure_of_XL_geometry}, becomes
\ber
[\bar\delta^{(+)}_1, \bar{\delta}^{(+)}_2]\bbX{\ell}
&=& \epb{+}_{[2}\epb{+}_{1]}\left[\mathcal{N}(\Up)^\ell_{jk}\bbDB{+}\bbX{j}\bbDB{+}\bbX{k}\right]\nn\\
&=&\epb{+}_{[2}\epb{+}_{1]}\bigr[(f_{\bar r,\ell}f_{r}-f_{r, \ell}f_{\bar r}+f_{\bar r,r}\tilde g-f_{r,\bar r}\bar{\tilde h})\bbDB{+}\bbX{r}\bbDB{+}\bbX{\bar r}\nn\\
&&
 +(f_{r,\ell}f_{\bar\ell}-f_{\bar\ell, \ell}f_r-f_{\bar\ell,r}\tilde g)\bbDB{+}\bbX{\ell}\bbDB{+}\bbX{r}
 +(f_{\bar r,\ell}f_{\bar\ell}-f_{\bar\ell, \ell}f_{\bar r}-f_{\bar\ell,\bar r}\bar{\tilde h})\bbDB{+}\bbX{\ell}\bbDB{+}\bbX{\bar r}\bigr]\nn\\
&=&\epb{+}_{[2}\epb{+}_{1]}f_{\bar \ell}\,\frac{K_{r\bar r \ell}}{K_{\ell\bar \ell}}\Bigr(\frac{2\tilde g}{c-1}\Bigr)_{,\ell}\bbDB{+}\bbX{r}\bbDB{+}\bbX{\bar r}\nn\\
&=& 0,
\eer{closure_of_barplus_barplus}
where in the last line we used the fact that $\tilde g$ and $c$ are constants. The vanishing of the Nijenhuis tensor for an almost complex structure means that the structure is integrable, hence a complex structure. Here we see that the relevant parts of the Nijenhuis tensor for the transformation matrix $\Up$ vanish. The same is true for the relevant parts of the Nijenhuis tensor for $\Um$ and $\Vpm$.

We now move over to the terms in the algebra \re{closure_of_XL_geometry} involving the Magri-Morosi concomitant. 
For clarity, we define the following combinations of the parameter functions,
\beq
\begin{array}{lcllcl}
	\mu &=& f_{\bar \ell} \bar{f}_r + f_r \tilde h, & \nu &=& f_{\bar \ell}\bar{f}_{\bar r} + f_{\bar r}\bar{\tilde g}~,\\
	\tau &=&f_{\bar \ell}(g-\bar h)-f_{r}\tilde f_{\bar\ell},\qquad
	&\omega &=& f_{\bar r} g-\tilde f_{\bar r} f_r~.
\end{array}
\eeq{define_mu_nu}
To investigate on-shell closure of the $(+)$-supersymmetry for $\bbX{\ell}$, we use the conjugate version of the field equation \re{7} to solve for $\bbD{+}\bbX{r}$. The algebra becomes
\ber
[\delta^{(+)}_1, \bar\delta^{(+)}_2]\bbX{\ell} &=&   \bar\epsilon^+_{[2}\epsilon^+_{1]} \left[\mathcal{M}(\Up, \Vp)^\ell_{jk}\bbDB{+}\bbX{j}\bbD{+}\bbX{k}+(\Up\Vp)^{\ell}_j\bbDB{+}\bbD{+}\bbX{j}\right]\nn\\
&=& \bar\epsilon^+_{[2}\epsilon^+_{1]} \,\bbDB{+}\left(-|f_{\bar\ell}|^2\bbD{+}\bbX{\ell}-\mu \bbD{+}\bbX{r}-\nu \bbD{+}\bbX{\bar r}\right)\nn\\
&=&  \bar\epsilon^+_{[2}\epsilon^+_{1]}\, \bbDB{+}\left(\Bigr(\mu \frac{K_{\ell\bar\ell}}{K_{\bar\ell r}}-|f_{\bar\ell}|^2\Bigr)\bbD{+}\bbX{\ell} + \Bigr(\mu\frac{K_{\bar\ell \bar r}}{K_{\bar \ell r}}-\nu\Bigr) \bbD{+}\bbX{\bar r}\right)\nn\\
&=&  \bar\epsilon^+_{[2}\epsilon^+_{1]} \, \bbDB{+}\left(\Bigr(|f_{\bar\ell}|^2-\tilde g\tilde h-|f_{\bar\ell}|^2\Bigr)\bbD{+}\bbX{\ell} + \frac{K_{\bar l\bar r}}{K_{\ell\bar\ell}}(\bar{\tilde g} \bar{\tilde h}-\tilde g\tilde h) \bbD{+}\bbX{\bar r}\right)\nn\\
&=& \bar\epsilon^+_{[2}\epsilon^+_{1]} i\pa{\+}\bbX{\ell}~,
\eer{closing_plus_part}
where in the last line we used that $\tilde g\tilde h=-1$. We already know that the $(-)$-part of the algebra for $\bbX{\ell}$ closes to a supersymmetry  if and only if $\tilde g$ and $\tilde h$ satisfies this constraint. The vanishing of the Magri-Morosi concomitant for two commuting complex structures is equivalent to the statement that the structures are simultaneous integrable \cite{Howe:1988cj}. Here we see that the relevant parts of the Magri-Morosi concomitant of $\Up$, $\Vp$ combines with products of the transformation matrices to vanish on-shell, such that the algebra closes to a supersymmetry.

Now we focus on the mixed $\epb{+}\epb{-}$-terms in the algebra. Again using the field equations \re{7} to write $\bbDB{-}{\bbX{\bar r}}$ in terms of  $\bbDB{-}\bbX{\ell}$ and $\bbDB{-}\bbX{\bar\ell}$, together with the expressions for the transformation functions in  \re{PDEs_original_form}-\re{PDEs_original_form_tilde}, the algebra closes to
\ber
[\bar\delta^{(+)}_1, \bar\delta^{(-)}_2]\bbX{\ell} 
&=& \bar\epsilon^+_{[2}\epb{-}_{1]}\left[\mathcal{M}(\Up, \Um)^\ell_{jk}\bbDB{+}\bbX{j}\bbDB{-}\bbX{k}-[\Up,\Um]^{\ell}_j\bbDB{+}\bbDB{-}\bbX{j}\right]\nn\\
&=& \bar\epsilon^+_{[2}\bar\epsilon^-_{1]} \, \bbDB{+}\left[\tau\bbDB{-}\bbX{\bar\ell}+\omega\bbDB{-}\bbX{\bar r}+(-f_r\tilde f_{\ell})\bbDB{-}\bbX{\ell}\right]\nn\\
&=& \bar\epsilon^+_{[2}\bar\epsilon^-_{1]} \, \bbDB{+}\left[\Bigr(\tau-\frac{K_{\bar \ell r}}{K_{r\bar r}}\omega\Bigr)\bbDB{-}\bbX{\bar\ell}+\Bigr(-f_r\tilde f_{\ell}-\frac{K_{\ell r}}{K_{r\bar r}}\omega)\bbDB{-}\bbX{\ell}\right]\nn\\
&=& \bar\epsilon^+_{[2}\bar\epsilon^-_{1]}\, \bbDB{+}(g\tilde g-\tilde g \bar h)\bbDB{-}\bbX{\ell}\nn\\
&=& 0~,
\eer{closing_mixed_part}
where in the last line we use that $g$ and $\tilde g$ are constants. A similar derivation can be done for the $\epb{+}\ep{-}$-term in the algebra, which also vanishes when using the field equations. 
The closure of the algebra on the right semichiral field follows in exactly the same way.

As a summary, we see that the transformations defined in \re{defining_ansatz_2} close to a supersymmetry algebra on the semichiral fields on-shell,
\beq
[\delta_1, \delta_2]\bbX{i}=\epb{+}_{[2}\ep{+}_{1]}i\pa{\+}\bbX{i} + \epb{-}_{[2}\ep{-}_{1]}i\pa{=}\bbX{i}~,
\eeq{closes_on_XL}
and that the action is invariant under the same transformations, if and only if the transformation functions take the expressions in  \re{PDEs_original_form}-\re{g_h_information} and 
\beq
|g|^2 = |\tilde g|^2 = - \frac{K_{\ell\bar\ell} K_{r\bar r}-|K_{\ell r}|^2}{K_{\ell\bar\ell} K_{r\bar r}-|K_{\ell \bar r}|^2}
\eeq{g_h_information_2}
is a constant.

%======================================================================
\section{Hyperk\"ahler solutions}
\label{HK}
The four-dimensional target space geometry is hyperk\"ahler if $c$ in \re{106} is a constant with absolute value less than one. We see from \re{hat_c_in_g} that this is the case at hand. In this section we explore some additional properties of this hyperk\"ahler geometry.
We  notice that the structures in \re{su2}  are now integrable and give us an $SU(2)$ of complex structures, 
\beq
[J^{(A)},J^{(B)}]=\delta^{AB}+\epsilon^{ABC}J^{(C)}~.
\eeq{hyperkahler}

To describe the hyperk\"ahler (HK) geometry, a  generalized potential $K$  must satisfy \re{106} with constant $|c|<1$. An additional requirement is that the the determinant of the matrix $K_{LR}$ is nonvanishing \re{determinant_related_to_bc}.
The transformations functions are then found from (derivatives of) this $K$ as solutions of  \re{PDEs_original_form}-\re{g_h_information}.

As discussed in the appendix, there are many quadratic actions which satisfy \re{106} and \re{determinant_related_to_bc}, \eg
\beq
K=(\bbX{\ell}-\bbX{\bar\ell})(\bbX{r}-\bbX{\bar r})+(\bbX{r}+\bbX{\bar r}+\bbX{\ell})(\bbX{r}+\bbX{\bar r}+\bbX{\bar\ell})~.
\eeq{flatex}
Solutions to \re{PDEs_original_form}-\re{g_h_information} are easily found for this  $K$ since all the coefficients are constants. The supersymmetry transformations are linear. More on this in the appendix.

There are a number of nontrivial examples of HK geometries written in semichiral coordinates. In particular in \cite{Bogaerts:1999jc}, the relation between the K\"ahler potential and the generalized K\"ahler potential is discussed, and HK geometries in semichiral coordinates are generated from K\"ahler potentials with certain isometries amongst their chiral and twisted chiral coordinates (see also \cite{Crichigno:2011aa} and \cite{Dyckmanns:2011ts} for further discussions 
of semichiral formulations of hyperkahler geometries). The method is an adaption of the  Legendre transform construction of 
\cite{Lindstrom:1983rt}, \cite{Hitchin:1986ea}.  In brief, the construction yields a semichiral description of a four-dimensional HK manifold from any function $F(x,v,\bar v)$ which satisfies Laplace's equation
\beq
F_{xx}+F_{v\bar v}=0~.
\eeq{Lapl}
Here $x$ and $v$ correspond to certain expressions in the chiral and twisted coordinates that need not concern us here.
The T-duality between a chiral and twisted chiral model with $N=(4,4)$ supersymmetry and its semichiral dual counterpart is investigated in detail in \cite{malin}.
The generalized K\"ahler potential is obtained via the Legendre transform
\ber\nn
&&K(\bbX{\ell}-\bbX{\bar \ell}, \bbX{\ell}+\bbX{\bar \ell}+2\bbX{r}, \bbX{\ell}+\bbX{\bar \ell}+2\bbX{\bar r})\\
&&=
F(x,v,\bar v)-\half v (\bbX{\ell}+\bbX{\bar \ell}+2\bbX{r})-\half \bar v (\bbX{\ell}+\bbX{\bar \ell}+2\bbX{\bar r})
-i\half x(\bbX{\ell}-\bbX{\bar \ell})~,
\eer{Ltf}
with
\ber\nn
F_x&=&i\half (\bbX{\ell}-\bbX{\bar \ell})=:i\half z~,\\\nn
F_v&=&\half (\bbX{\ell}+\bbX{\bar \ell}+2\bbX{ r})=:\half y~,\\
F_{\bar v}&=&\half  (\bbX{\ell}+\bbX{\bar \ell}+2\bbX{\bar r})=:\half \bar y~.
\eer{Fder}
The resulting $K$ satisfies \re{106} with\footnote{The  restriction to $c=0$ is not necessary for this kind of construction--see \cite{Crichigno:2011aa}.} $c=0$ and has $\hbox{det}(K_{LR})\ne 0$.

We now use this construction to generate a nontrivial example illustrating the discussion in the previous sections. Since we do not need the connection to a K\"ahler potential, we can start from any $F$ satisfying \re{Lapl}. A convenient example is\footnote{When $v$ and $x$ are chiral and real linear superfields, respectively, $F$ is the superspace Lagrangian of the improved tensor multiplet \cite{Karlhede:1984vr}.}
\beq
F(x,v,\bar v)=r-x \hbox{ln}(x+r)~, ~~~~~r^2:=x^2+4v\bar v~.
\eeq{EX}
Solving the relations \re{Fder} and plugging into \re{Ltf} results in the generalized K\"ahler potential
\beq
K=\half e^{-\half i z}\left( 1-{\textstyle{\frac 1 4}y\bar y}\right)~,
\eeq{GKex}
which indeed satisfies all the relevant requirements. Note that it is written in coordinates invariant under the Abelian symmetry
\beq
\delta \bbX{\ell} =\varepsilon~, \quad \delta \bbX{r} =-\varepsilon~, ~~\varepsilon \in \mathbb{R}~.
\eeq{abelian}
To find the transformation functions, we calculate the various second derivatives of  $K$ and insert into
\re{PDEs_original_form}-\re{g_h_information}. The resulting partial differential equations may then be solved to yield
\ber\nn
f&=&2i(\lambda-\tilde g ) \hbox{ln} (2-iy)+2i(\lambda+\tilde g )\hbox{ln}(2-i\bar y)+\lambda z~,\\
\tilde f&=&-i\kappa \,\hbox{ln}(\bar y)+ ig{\textstyle {\frac 1 8}}y^2-\half(\kappa+g)z~,
\eer{atlast}
where $\lambda$ and $\kappa$ are integration constants. The appearance of these constants  may seem surprising, since we expect the transformations to be unique. Below we shall see how they are determined. 

As in the derivation of the supersymmetry 
transformations in \cite{Lindstrom:2004eh}, we identify part of  \re{atlast} as  field equation symmetries, that is symmetries of a Lagrangian ${\cal L}(\varphi)$ of the form
\beq
\delta \varphi^i=A^{ij}\frac {\partial {\cal L}}{\partial\varphi^j}~,
\eeq{varphi}
with $A^{ij}$ anti-symmetric (or symmetric for spinorial indices). These transformations will leave the action invariant and vanish on-shell.
When inserted into \re{defining_ansatz_2}  the $\lambda$ part of $f$ gives an expression that vanishes due to the $\bbX{\ell}$ field equation,
\beq
(2-iy)(2-i\bar y)\bbDB{+}z+2(2-i\bar y)\bbDB{+}y+2(2-iy)\bbDB{+}\bar y=0~,
\eeq{reduceconstants1}
and the $\kappa$ part of $\tilde f$ gives an expression that vanishes due to the $\bbX{r}$ field equation,
\beq
\bar y\bbDB{-}z+2i\bbDB{-}\bar y=0~.
\eeq{reduceconstants2}
This means that the transformations \re{atlast} reduce to
\ber\nn
f&=&2i\tilde g (\,\hbox{ln} (2-iy) + \hbox{ln}(2-i\bar y))~,\\
\tilde f&=& g(i{\textstyle {\frac 1 8}}y^2-\half z)~.
\eer{atlast2}
According to \re{g_h_information_2}, $\tilde g$ and $g$ are phases when $c=0$, and from \re{defining_ansatz_2} we see that they may be absorbed by an R-transformation of the charges $Q_\pm$ for the extra supersymmetries :
\ber\nn
&&\tilde g =: e^{i\varphi}~,  \quad g =:e^{i\psi}~,\\
&&\bar\epsilon ^+ e^{i\varphi}  \to \bar\epsilon ^+~, \quad \bar\epsilon ^- e^{i\psi}  \to \bar\epsilon ^-~.
\eer{reparametrization}
The final form of the functions $f$ and $\tilde f$ thus becomes
\ber\nn
f&=&2i\,\hbox{ln}\left(\frac {2-i\bar y}{2-iy}\right)~,\\
\tilde f&=&i\half \left(iz+{\textstyle{\frac 1 4 }}y^2\right)~.
\eer{finalf}
Inserting this in \re{defining_ansatz_2} yields the transformations
\ber\nn
\delta \bbX{\ell} &=&-\frac{\bar\epsilon^+ }{(2-iy) (2-i\bar y)}2\left(i(y-\bar y)\bbDB{+}\bbX{\bar\ell}+2(2-i\bar y)\bbDB{+}\bbX{r}-2(2-iy)\bbDB{+}\bbX{\bar r}\right)\\[2mm]\nn&&+\bar\epsilon^- \bbDB{-}\bbX{\ell}-\epsilon^- \bbD{-}\bbX{\ell}~,\\[2mm]
\delta \bbX{r}&=&-\frac{\bar\epsilon^-} 4\left((2-iy)\bbDB{-}\bbX{\ell}-(2+iy)\bbDB{-}\bbX{\bar\ell}\right)+\bar\epsilon^+ \bbDB{+}\bbX{r}-\epsilon^+\bbD{+}\bbX{r}~,
\eer{transfagain}
and their complex conjugates.

%======================================================================
%======================================================================
\section{Summary and conclusions}
\label{summary}

We have found that our ansatz \re{defining_ansatz_2} for additional supersymmetries of a semichiral sigma model with four-dimensional target space corresponds to hyperk\"ahler geometry on  the target space. For this case we have provided the form of the transformation functions, related them to previous general discussions in the literature and described generalized K\"ahler potentials satisfying the invariance conditions. 

The existence of four-dimensional examples of semichiral sigma models with non-trivial $B$-field \cite{S1xS3} indicates that the ansatz \re{defining_ansatz_2} has to be modified for on-shell algebras. It is well known, \eg from four-dimensional sigma models with chiral fields $(\phi,\bar \phi)$ as coordinates, that extra supersymmetries may come together with  central charge transformations \cite{Lindstrom:1983rt}, \cite{Hull:1985pq} in the form
\beq
\delta \phi^i=\bar D^2(\epsilon\Omega^i )~,
\eeq{cch}
where the scalar transformation superfield $\epsilon$ contains the supersymmetry at the $\theta$ level. Central charge transformations vanish on-shell, but will have effect, e.g.,  on the conditions that follow from invariance of the action. Such a generalization of the transformations 
\re{defining_ansatz_2} will presumably cover the $dB\ne0$ case.

\bigskip\bigskip
\noindent{\bf\Large Acknowledgement}:
\bigskip\bigskip

\noindent

The research of UL was supported by  VR grant 621-2009-4066. MR acknowledges NSF Grant phy 0969739.
\appendix
\section{Linear transformations}

In \cite{Goteman:2009xb} we were able to extract interesting information from a semichiral sigma model with four-dimensional target space and additional non-manifest linear {\it pseudo}-supersymmetry.  In this section we show that linear supersymmetry transformations are only possible for quadratic potentials (flat geometry) or  when the metric is degenerate.

There is a very direct argument why linear transformations lead to quadratic actions. 
If  the transformation on a field $X$ is linear and the algebra closes on-shell, we have, schematically,
\beq
\delta X=QX,~~\Rightarrow Q^2X=[\delta,\delta]X=\partial X+F
\eeq{}
where $F$ is (derivatives of) a field equation. The latter then reads
\beq
F={\cal O}X=(Q^2-\partial)X=0~,
\eeq{}
which means that the Lagrangian must be quadratic.  In the present case, it still turns out to be instructive to explicitly consider the linear case. 

 The transformations are the same as in the general case \re{defining_ansatz_2}, with the difference that the transformation functions are now all constant.
Again, off-shell closure of the algebra cannot occur, since $|f_\ell|^2\neq -1$. On-shell closure, however, can be obtained just as in the general case.

Invariance of the action \re{action} under the linear transformations implies the same system of partial differential equations \re{PDEs_original_form}-\re{PDEs_original_form_tilde}, but with constant parameters. The equations imply that the Lagrangian $K$ must satisfy
\beq
	K_{\ell r} = \frac{\bar\nu}{\mu} K_{\bar \ell r}~, 
\eeq{all_PDEs_linear}
among other relations. The parameters $\mu$ and $\nu$ are defined in \re{define_mu_nu} and are here constant. Taking derivative with respect to $\bbX{r}$ on both sides implies that $(|\mu|^2-|\nu|^2) K_{\ell r \bar r}=0$. This implies that either $|\mu|^2=|\nu|^2$, leading to degenerate metric, as we have seen in \re{determinant_related_to_bc}, or that $K_{\ell r\bar r}=0$. Similar results can be derived from the other PDEs and as a result we draw the conclusions that linear supersymmetry transformations imply either degenerate metric, or a generalized K\"ahler potential with vanishing third derivatives, i.e. a quadratic potential.

For a quadratic potential $K$, everything works as in the general case. From the  equation \re{106} 
we  again find that
\beq
c=\frac{1-|g|^2}{1+|g|^2} \quad \Longleftrightarrow \quad |g|^2=\frac{1-c}{1+c}~.
\eeq{18}
This again implies that $|c|<1$ and thus the geometry is necessarily  hyperk\"ahler.

As a final comment we note that there are a large set of non-trivial (non-quadratic) generalized potentials invariant under the linear transformations with $\hbox{det}(K_{LR})=0$. As mentioned, this makes it impossible to extract a metric, so they do not correspond to sigma models. One may speculate that these have an application in models where the background is in some sense topological.

%======================================================================
%======================================================================

\section{Complex structures}
In \cite{Goteman:2009ye} it is concluded on general grounds that the  relation \re{onshellcndn} holds  on-shell
\beq
\half \alpha~\!\left(J^{(1)}_{(\pm)}-iJ^{(2)}_{(\pm)}\right)=U^{(\pm)}\pi^{(\pm)}~,
\eeq{}
where 
\beq
\pi^{(\pm)}:=\half (\one +iJ^{(\pm)})~,
\eeq{pidefinition}
 $J^{(A)}_{(\pm)}$ are the complex structures obeying an $SU(2)$ algebra with $J^{(3)}_{(\pm)}:=\Jpm$ and $\alpha$ is a phase\footnote{This phase is not included in \cite{Goteman:2009ye}, but represents an ambiguity in the choice of $\epsilon$ related to $R$-symmetry.}. Here we verify this relation by explicitly constructing $J^{(1)}$ and $J^{(2)}$ as in \re{su2}.

From \cite{Lindstrom:2005zr}, 
$\Jpm$  takes the form
\ber\nn
\Jp &=&\frac i D\left(\begin{array}{cccc}
D&0&0&0 \cr
0&-D&0& 0\cr
2K_{\ell \bar\ell}K_{\ell \bar r}&2K_{\ell \bar\ell}K_{\bar\ell\bar r}&S&2 K_{\ell \bar r} K_{\bar\ell \bar r} \cr
-2 K_{\ell \bar\ell}K_{\ell r}&-2 K_{\ell \bar\ell} K_{\bar\ell r}&-2K_{\ell r}K_{\bar\ell  r}&- S\end{array}\right),\\[2mm]
\Jm&=&\frac i D\left(\begin{array}{cccc}
S&2K_{\bar\ell\bar r}K_{\bar\ell r}&2K_{r \bar r}K_{\bar\ell r}&2K_{r \bar r}K_{\bar\ell \bar r}\cr
-2K_{\ell r}K_{\ell \bar r}&-S&-2K_{r \bar r}K_{\ell  r}& -2K_{r \bar r}K_{\ell \bar r}\cr
0&0&D&0  \cr
0&0&0&-D\end{array}\right).
\eer{1099}
where we have defined the sum and difference
\beq
S:= |K_{\ell r}|^2+|K_{\ell \bar r}|^2~, ~~~ D:= |K_{\ell r}|^2- |K_{\ell \bar r}|^2~.
\eeq{}
Inserting the expressions for $\Jpm$ in \re{su2} yields
\beq
J^{(1)}=\frac {2i} {D\sqrt{1-c^2}}\left(\begin{array}{cccc}
K_{\ell \bar\ell}K_{r \bar r}&K_{\bar\ell r}K_{\bar\ell\bar r}&K_{r \bar r}K_{\bar\ell  r}& K_{r \bar r}K_{\bar\ell \bar r}\cr
-K_{\ell \bar r}K_{\ell r}&-K_{\ell \bar\ell}K_{r \bar r}& -K_{r \bar r}K_{\ell r}& -K_{r \bar r}K_{\ell \bar r}\cr
c K_{\ell \bar\ell}K_{\ell \bar r}&c K_{\ell \bar\ell}K_{\bar\ell\bar r}&\half (D+cS)&c K_{\bar\ell\bar r} K_{\ell \bar r}\cr
-c K_{\ell \bar\ell}K_{\ell r}&-c K_{\ell \bar\ell} K_{\bar\ell  r}&-c  K_{\ell r}K_{\bar\ell  r}&-\half(D+c S)\end{array}\right)~,
\eeq{109}
and
\beq
J^{(2)}=-\frac {2}{D \sqrt{1-c^2}}
\left(\begin{array}{cccc}
K_{\ell \bar\ell}K_{r\bar r}&K_{\bar\ell r} K_{\bar\ell\bar r}&K_{r\bar r}K_{\bar\ell r}& K_{r\bar r}K_{\bar\ell\bar r}\cr
K_{\ell \bar r}K_{\ell r}&K_{\ell \bar\ell}K_{r \bar r}&K_{r \bar r}K_{\ell r}& K_{r \bar r}K_{\ell \bar r}\cr
-K_{\ell \bar\ell}K_{\ell \bar r}&-K_{\ell \bar\ell}K_{\bar\ell\bar r}&-K_{\ell \bar\ell}K_{r \bar r}& -K_{\bar\ell \bar r} K_{\ell \bar r}\cr
-K_{\ell \bar\ell}K_{\ell r}&-K_{\ell \bar\ell}K_{\bar\ell r}& -K_{\ell r}K_{\bar\ell r}&-K_{\ell \bar\ell}K_{r \bar r}\end{array}\right).
\eeq{10999}

Finally, we have
\ber\nn
D \sqrt{1-c^2}~\Up \pip=
\eer{}
\ber\nn
\left(\begin{array}{cccc}
(f_{\bar r}K_{\ell r}-f_rK_{\ell \bar r})K_{\ell \bar\ell}&f_{\bar\ell}D-(f_rK_{\bar\ell\bar r}-f_{\bar r}K_{\bar\ell  r})K_{\ell \bar\ell}&  (f_{\bar r}K_{\ell r}-f_r K_{\ell \bar r})K_{\bar\ell  r}& (f_{\bar r}K_{\ell r}-f_rK_{\ell \bar r})K_{\bar\ell\bar r}\cr
0&0&0&0\cr
-\tilde gK_{\ell \bar\ell}K_{\ell \bar r}&-\tilde g K_{\ell \bar \ell}K_{\bar\ell \bar r}& -\tilde g|K_{\ell \bar r}|^2 & ~-\tilde gK_{\bar\ell\bar r} K_{\ell \bar r}\cr
\bar{\tilde h} K_{\ell \bar\ell}K_{\ell r}&\bar{\tilde h}K_{\ell \bar\ell}K_{\bar\ell  r}& \bar{\tilde h}K_{\ell  r}K_{\bar\ell  r}& \bar{\tilde h}|K_{\ell r}|^2\end{array}\right).
\eer{}
\ber
~
\eer{11992}
Using these expressions for the  HK models at hand, we verify explicitly that  \re{onshellcndn} is satisfied with $i\alpha =\tilde g=-\bar{\tilde h}$ due to the invariance conditions \re{PDEs_original_form}-\re{g_h_information}.

\section{Example metric}

The metric\footnote{In a semichiral model with four-dimensional target space, the (trivial) $B$-field, is proportional to $c$ for a HK metric. Since $c=0$ in the example, it should vanish. The formulae in \cite{Lindstrom:2005zr} confirm this.} that follows from the potential \re{GKex} may be found from the formulae in \cite{Lindstrom:2005zr}. It is
\beq
g=\left(\begin{array}{cc}
g_{LL}&g_{LR}\cr
g_{RL}&g_{RR}\end{array}\right)
=\frac {ie^{-i\half z}}{2(y-\bar y)}\left(\begin{array}{cc}
\mathbb{A}&\mathbb{B}\cr
\mathbb{B}^t&\mathbb{C}\end{array}\right)
\eeq{}
with
\ber\nn
\mathbb{A}&=&\textstyle{\frac 1 {16}}(4+y\bar y)\left(\begin{array}{cc}
4-y\bar y-2i(y+\bar y)&4+y\bar y\cr
4+y\bar y&4-y\bar y+2i(y+\bar y)\end{array}\right)\\[2mm]\nn
\mathbb{B}&=&\textstyle{\frac 1 {4}}\left(\begin{array}{cc}
(2-i\bar y)[4+y\bar y-i(y+\bar y)]&(2-i y)[4+y\bar y+i(y+\bar y)]\cr
(2+i\bar y)[4+y\bar y-i(y+\bar y)]&(2+i y)[4+y\bar y+i(y+\bar y)]\end{array}\right)\\[2mm]
\mathbb{C}&=&\left(\begin{array}{cc}
4+\bar y^2&4+y\bar y\cr
4+y\bar y &4+y^2\end{array}\right)
\eer{}
\eject

\end{document}